\documentclass[aps,prb,twocolumn,showpacs,amsmath,amssymb,superscriptaddress,10pt]{revtex4-1}

\usepackage[pdftex]{graphicx}
\usepackage[colorlinks=true,citecolor=blue]{hyperref}
\usepackage{braket}
  
\newcommand{\up}{\uparrow}
\newcommand{\down}{\downarrow}
\newcommand{\tL}{\text{L}}
\newcommand{\tR}{\text{R}}
\newcommand{\tx}{\text}

\renewcommand{\vec}[1]{\ensuremath{\mathbf{#1}}}

\newcommand{\Gm}{\Gamma}
\newcommand{\sg}{\sigma}

\newcommand{\Abs}[1]{\ensuremath{\left| #1 \right|}}

\newcommand{\kB} {k_\text{B}}
\newcommand{\w}{w}

\begin{document}
\title{Coherent dynamics in stochastic systems revealed by full counting statistics}
 
\author{Philipp Stegmann}\email{philipp.stegmann@uni-due.de}
\affiliation{Theoretische Physik, Universit\"at Duisburg-Essen and CENIDE, 47048 Duisburg, Germany}

\author{J\"urgen K\"onig}
\affiliation{Theoretische Physik, Universit\"at Duisburg-Essen and CENIDE, 47048 Duisburg, Germany}

\author{Stephan Weiss}\email{weiss@thp.uni-due.de}
\affiliation{Theoretische Physik, Universit\"at Duisburg-Essen and CENIDE, 47048 Duisburg, Germany}

\date{\today}

\begin{abstract}
Stochastic systems feature, in general, both coherent dynamics and incoherent transitions between different states.
We propose a method to identify the coherent part in the full counting statistics for the transitions.
The proposal is illustrated for electron transfer through a quantum-dot spin valve, which combines quantum-coherent spin precession with electron tunneling.
We show that by counting the number of transferred electrons as a function of time, it is possible to distill out the coherent dynamics from the counting statistics even in transport regimes, 
in which other tools such as the frequency-dependent current noise and the waiting-time distribution fail.
\end{abstract}


\maketitle

% - - - - - - - - - - - - - - - - - - - - - - - - - - - - - - - - - - - - - -
\section{Introduction}
In a stochastic process, the evolution of a system is described in terms of random events~\cite{kampen_stochastic_2011}.
A generic example is the tunneling of electrons into and out of quantum dots coupled to electron reservoirs.
During the time between two tunneling events, the quantum-dot state undergoes a quantum-coherent evolution.
A fast coherent evolution (as compared to the rate of tunneling) may easily dominate the overall dynamics of the system.
In the opposite limit, the probabilistic nature of incoherent tunneling prevails and it may seem a hopeless task to detect features of the coherent dynamics by just counting the number of tunneled electrons as a function of time.
In this paper, however, we propose a method based on full counting statistics to distill out the contributions stemming from coherent evolution. 

\begin{figure}[ht!]
	\includegraphics[width=\columnwidth]{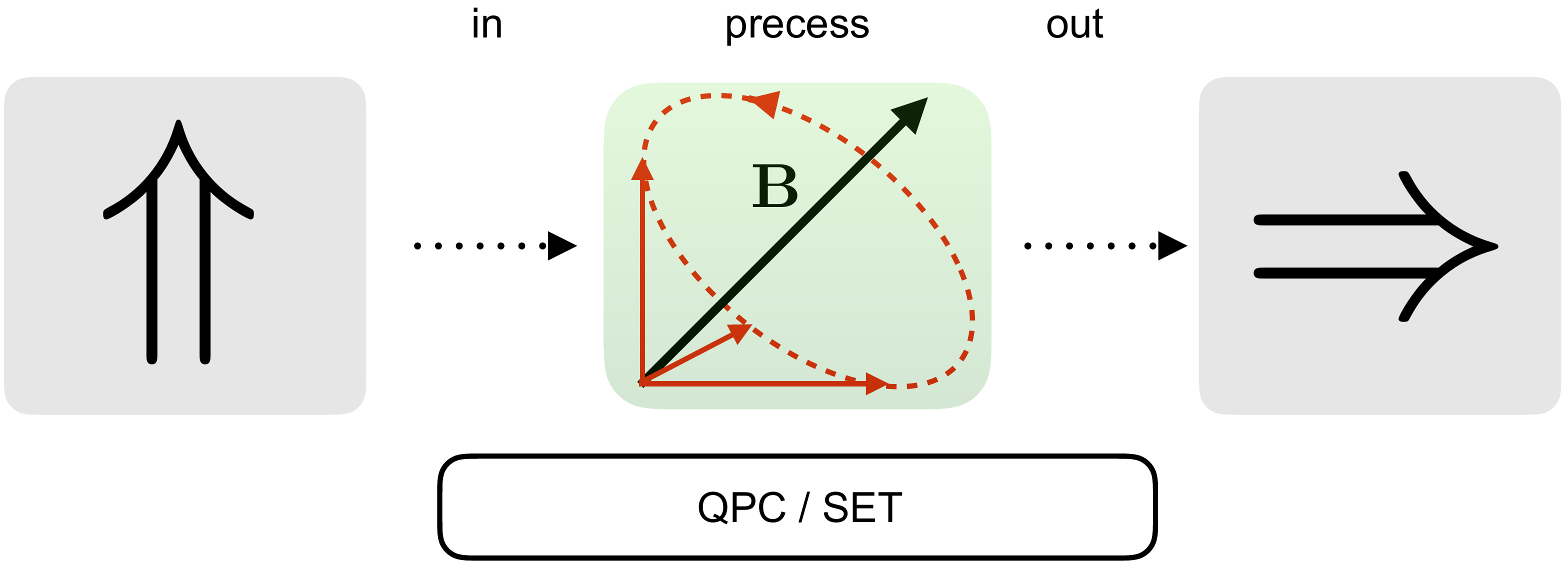}
	\caption{\label{fig:model}(Color online) Electron transfer through a quantum-dot spin valve: an electron tunnels in from the left lead, precesses about the exchange field $\mathbf{B}$, and then tunnels out to the right lead. A coupled quantum-point contact (QPC) or single-electron transistor (SET) measures the electron occupation of the quantum dot.
}
\label{fig1}
\end{figure}

To illustrate our proposal we choose as an example a quantum-dot spin valve (see Fig.~\ref{fig:model}).
It consists of a single-level quantum dot attached to two ferromagnetic leads with non-collinear magnetization directions.
Quantum-dot spin valves have been realized experimentally with metallic nanoparticles~\cite{deshmukh_using_2002,davidovic_2005,bernand_anisotropic_2009}, semiconductor quantum dots~\cite{hamayal_spin_2009} and molecules~\cite{yoshida_gate_2013} as well as in InAs nanowires~\cite{hofstetter_ferromagnetic_2010} and carbon nanotubes~\cite{sahoo_electric_2005,crisan_harnessing_2016}. 
An applied bias voltage yields a finite polarization of the quantum-dot spin. 
The coupling of the quantum-dot level to ferromagnetic leads generates an exchange field that gives rise to a coherent Larmor precession of the accumulated spin~\cite{koenig_interaction_2003, braun_theory_2004}. 
The relative orientation of the quantum-dot spin and the magnetization of the drain electrode affects the probability for the electron to tunnel out. 
Therefore, the coherent spin dynamics influences the incoherent tunneling transport.
A time-resolved monitoring of the individual tunneling events can be achieved by electrostatically coupling the quantum dot to a quantum point contact~\cite{gustavsson_counting_2005, gustavsson_electron_2009, fujisawa_bidirectional_2006, gustavsson_measurements_2007, fricke_bimodal_2007, flindt_universal_2009} or a single-electron transistor~\cite{lu_real_2003, bylander_current_2005}.
Such a coupling is spin insensitive and does, therefore, not affect the coherent spin dynamics.

Suppose that transport follows a fully deterministic cycle in which a majority-spin electron enters from the source electrode into the quantum dot, precesses with Larmor frequency in the exchange field by an azimuthal angle of $\pi$ (such that the relative angle between the quantum-dot spin and the majority-spin direction of the drain electrode is minimized), leaves to the drain electrode, and, thereafter, immediately the next electron enters from the source.
Consequently, the coherent dynamics would be directly visible in the sequence of equidistant charge-transfer events occurring with Larmor frequency.
The probabilistic nature of incoherent tunneling events, however, destroys this regular pattern.
One reason is that the instant of time at which the precession starts depends on how long the quantum dot remains empty before an electron tunnels in.
This disturbing factor can be eliminated by correlating tunneling events to each other, e.g., by studying either the waiting-time distribution~\cite{brandes_waiting_2008,rajabi_waiting_2013,sothmann_electronic_2014,potanina_electron_2017,tang_spin_2018,walldorf_electron_2018} or the frequency-dependent current-current correlator~\cite{braun_frequency_2006,sothmann_transport_2010}.
Both methods, however, still suffer from the fact that with some probability also minority spins may tunnel in and, furthermore, that tunneling out occurs also for non-optimal angles between quantum-dot spin and drain magnetization direction.
The reliability of both methods is restricted to relatively strong lead polarizations for resolving spin precession in quantum-dot spin valves.

In this paper, we propose an approach that is qualitatively different from analyzing correlators.
Instead, we suggest to simply average over the number $N$ of transferred electrons in a time interval of length $t$, including a weighting factor $s^N$ for the measured transfer probabilities $P_N(t)$,  
\begin{equation}
 	\langle N \rangle_s (t) := \frac{\sum\limits_{N=0}^{\infty} N s^N P_N(t)}{\sum\limits_{N=0}^{\infty} s^N P_N(t)} \, .
	\label{genN}
\end{equation}
Note that $\langle N \rangle_1(t)$ is the total number of transferred electrons. It increases linearly in time and displays no signatures of spin precession. For $s>0$, $\langle N \rangle_s (t)$ has been successfully applied to study dynamical phase transitions, e.g., in structural glass formers~\cite{hedges_dynamic_2009} or optical systems~\cite{garrahan_thermodynamics_2010}.

The weighting factor $s^N$ introduced in Eq.~\eqref{genN} has dramatic consequences for $s<0$. 
Then, a regular pattern of divergencies in $\langle N \rangle_s (t)$, separated in time by $\pi/ \Omega$, which is approximately half of the period of the spin precession, is observed.
It, thus, appears that the weighting factor tends to distill out the contributions relevant for the coherent dynamics.
Quantitatively, this new method works for a much larger parameter range than the analysis of waiting times or current-current correlators.
In practice, it remains to measure the probability distribution with sufficient precision as discussed in Sec.~\ref{sec:req}.

\section{Model \& Method}\label{sec:model}
The quantum-dot spin valve is described by the Hamiltonian $\mathcal{H}=\mathcal{H}_\text{dot}+\sum_{r=\tx{L,R}}\mathcal{H}_r+\mathcal{H}_{\text{tun}}$. 
The quantum dot, $\mathcal{H}_\text{dot}=\epsilon \sum_{\sigma} n_\sigma +Un_\uparrow n_\downarrow$, hosts a single, spinful level. 
Its energy $\epsilon$ can be tuned by a gate voltage.
The corresponding number operator is $n_\sigma=d_\sigma^\dag d_\sigma$ where the fermionic operator $d_\sigma^\dag$ ($d_\sigma$) creates (annihilates) an electron with spin $\sigma$ (with respect to some arbitrarily chosen spin-quantization axis). 
The charging energy for double occupation of the quantum dot is denoted by $U$.
The ferromagnetic leads are described as reservoirs of noninteracting electrons $\mathcal{H}_r=\sum_{\vec k \sigma} \varepsilon_{\vec k \sigma} a^{\dagger}_{r \vec k \sigma} a_{r \vec k \sigma},$ held at electrochemical potential $\mu_\tx{L}= eV/2$ and $\mu_\tx{R}= -eV/2$. 
Here, the spin-quantization axes are chosen along the respective magnetization direction $\vec n _r$ (enclosing an angle $\phi$) such that the operator $a^{\dagger}_{r \vec k \sigma}$ ($a_{r \vec k \sigma}$) creates (annihilates) an electron with momentum $\vec k$ and majority ($\sg = +$) or minority ($\sg = -$) spin. 
The degree of spin polarization $p_r=(\nu _{r +}-\nu _{r -})/(\nu _{r +}+\nu _{r -})$ of lead $r$ is characterized by the spin-dependent density of states~$\nu _{r \sigma}$ taken at the Fermi energy.
In the following, we assume that the leads are made of the same material such that $p_\tx{L}=p_\tx{R}=p$. 

The tunneling Hamiltonian reads as $\mathcal{H}_{\tx{tun}}=\sum_{r \vec k  \sigma \sigma'}t_{r\sigma \sigma'} d_{\sigma}^{\dagger} a_{r \vec k \sigma'} +\text{H.c.}$, with matrix elements $t_{r\sigma \sigma'} = t_r \langle \sigma | \sigma' \rangle_r$ that separate into the spin-independent bare tunneling amplitudes $t_r$ and the overlap factors $\langle \sigma | \sigma' \rangle_r$ accounting for different quantization axes in the quantum dot and lead $r$.
We define the tunnel-coupling strength $\Gamma_{r \pm}=2\pi | t_r|^2 \nu_{r \pm} =\left(1\pm p_r \right)\Gamma_r$ with $\Gamma_{r}= \left( \Gamma_{r +}+\Gamma_{r -}\right)/2$ as well as $\Gamma = \Gamma_\tx{L}+\Gamma_\tx{R}$. 
Finally, the asymmetry $a=(\Gamma_\tx{L}-\Gamma_\tx{R})/\Gamma$ measures the difference of the coupling strengths to the left and right leads.

We assume the dot level to be well inside the energy window provided by the transport voltage, $-eV/2 < \epsilon < eV/2$, and the energy to add a second electron outside, $eV/2< \epsilon+U$. 
At low temperature, $k_\tx{B} T \ll eV$, the quantum dot can, then, be either empty or singly occupied, and electron transport only occurs from the left lead through the dot to the right lead, while tunneling in the opposite direction can be neglected.

We calculate
\begin{equation}
 	\langle N \rangle_s (t) = s \frac{\partial \ln \mathcal{M}_s(z,t)}{\partial z} \bigg|_{z=0}
	\label{genNM}
\end{equation}
by making use of the generalized-factorial-moment generating function $\mathcal{M}_s(z,t)=\sum_{N=0}^\infty (z+s)^NP_N(t)$.
Following along the lines of Refs.~\onlinecite{stegmann_detection_2015,stegmann_short_2016}, the latter is calculated from $\mathcal{M}_s(z,t)={\mathbf e}^T \exp( {\mathbf W}_{\!z+s} t) {\boldsymbol{\rho}}_{\rm stat}$, where ${\boldsymbol{\rho}}_{\rm stat} = (\rho_{\rm stat}^{00},\rho_{\rm stat}^{\up\up},\rho_{\rm stat}^{\down\down},\rho_{\rm stat}^{\up\down},\rho_{\rm stat}^{\down\up})^T$ is the vector of matrix elements of the stationary quantum dot's reduced density matrix obtained from ${\mathbf W}_{1} {\boldsymbol{\rho}}_{\rm stat} = {\mathbf 0}$ and ${\mathbf e}^T  {\boldsymbol{\rho}}_{\rm stat} =1$ with ${\mathbf e}^T = (1,1,1,0,0)$. The explicit form of the generator $\mathbf{W}_{\!z}$ is given in Appendix~A.
The matrix elements $\rho_{\rm stat}^{00}$, $\rho_{\rm stat}^{\up\up}$, and $\rho_{\rm stat}^{\down\down}$ denote the probability to find the quantum dot empty or singly occupied with spin $\up$ or $\down$, respectively.
The remaining ones, $\rho_{\rm stat}^{\up\down} = \left( \rho_{\rm stat}^{\down\up} \right)^*$, describe coherent superpositions. 
Other matrix elements are exponentially suppressed in the considered transport regime.

The finite spin polarization $p$ of the leads enters ${\mathbf W}_{\!z}$ in two ways.
First, it affects the rate for tunneling in from the left and tunneling out to the right lead.
Second, it gives rise to an exchange field~\cite{koenig_interaction_2003,braun_theory_2004} that is (up to a factor $g\mu_\tx{B}$) given by
\begin{equation}
\label{eq:Bfield}
\mathbf{B}=\sum_r \frac{p \Gamma_r}{2\pi} \left[\Psi (\epsilon+U-\mu_r) -\Psi (\epsilon-\mu_r) \right] \mathbf{n}_r \,,
\end{equation}
where $\Psi (x)= \tx{Re} \, \psi (\frac{1}{2} + i \frac{x}{2 \pi k_\tx{B}T})$ is the real part of the digamma function $\psi$. 
The exchange field leads to the coherent precession of the quantum-dot spin that we want to detect by full counting statistics.

% - - - - - - - - - - - - - - - - - - - - - - - - - - - - - - - - - - - - - -
\section{Results}\label{sec:results}

\subsection{Full counting statistics}

\begin{figure}[t]
\center
\includegraphics[width=\columnwidth]{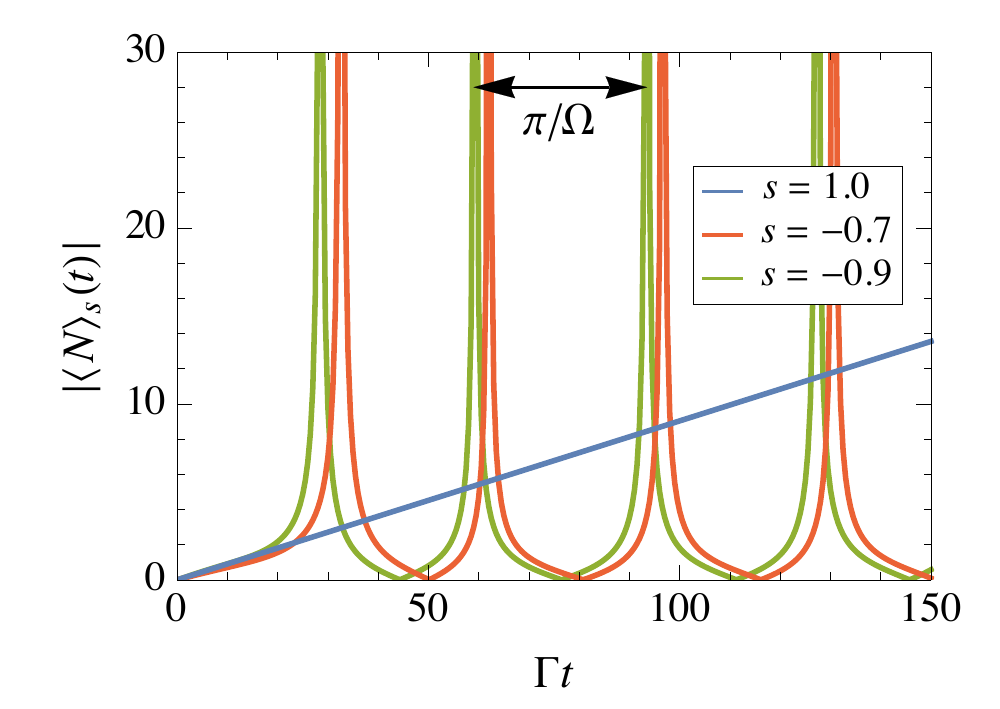}
\caption{\label{fig:fig2} (Color online) $|\langle N\rangle_s(t)|$ as a function of time for $s=1,-0.7,$ and $-0.9$. A regular peak pattern with peak-to-peak distance $\pi/\Omega$ occurs for negative $s$, where $\Omega$ is approximatively given by the Larmor frequency $|\mathbf{B}|$ of the exchange field. Parameters are $p=0.3$, $\phi=\pi/2$, $\epsilon=U/3$, $eV=13U/6$, $\kB T=U/30$, and $a=0.8$.
}
\end{figure}

In Fig.~\ref{fig:fig2}, we depict $|\langle N\rangle_s(t)|$ for $s=1$, $-0.7$, $-0.9$ and weakly-polarized leads, $p=0.3$ (as for Ni). Other parameters are $\epsilon=U/3, eV=13U/6, \kB T=U/30$. Counting the number of transferred electrons without weighting factor, $\langle N\rangle_1(t)$, trivially yields a linear time dependence (blue line).
The behavior of $|\langle N\rangle_s(t)|$ for values of $s<0$ is strikingly different.
It shows a periodic sequence of very sharp divergencies with a peak-to-peak distance $\pi/\Omega$ that is independent of $s$ [up to a term $\propto \mathcal{O}(p^3)$ that becomes also $s$-independent for $\Abs{s}\gg 1$].
This periodic pattern reflects the Larmor precession of the quantum-dot spin in the exchange field.
The strikingly clear signature is remarkable in at least two respects.
First, we note that the shown time interval allows for only $4$ Larmor precessions of angle $\pi$ each, but about $14$ electrons have been transferred through the quantum dot in total.
This means that the incoherent part of the dynamics dominates over the coherent part.
Nevertheless, the weighting factors are able to distill out the coherent evolution.
Second, we emphasize that $\langle N\rangle_s(t)$ is an average value over all possible realizations, in particular, over all initial quantum-dot states, i.e., it is {\it not} necessary to prepare the quantum dot in a specific initial state.
\begin{figure}[t]
\center
\includegraphics[width=\columnwidth]{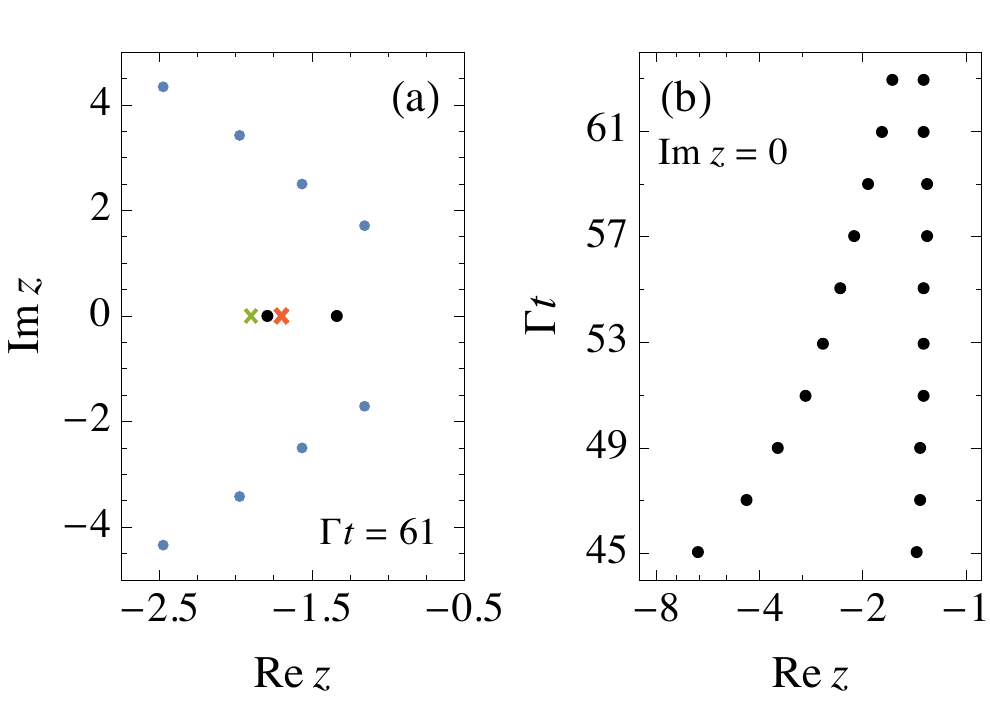}
\caption{\label{fig:fig3} (Color online) (a) Position of the zeros $z_j(t)$ of $\mathcal{M}_1(z,t)$ in the complex plane for $\Gamma t=61$.
Complex-valued zeros (blue dots) witness the presence of Coulomb interaction in the system. 
Real-valued zeros (black dots) indicate the presence of Larmor precession. $\langle N\rangle_{-0.7}$ and $\langle N\rangle_{-0.9}$ diverge when a real-valued zero crosses the points marked by a red and green cross, respectively.
(b) Position of the real-valued zeros (black dots) as a function of time. Other parameters are as in Fig.~\ref{fig:fig2}.
}
\end{figure}

The divergencies of $\langle N\rangle_s(t)$ are connected to the positions of the zeros $z_j(t)$ of $\mathcal{M}_1(z,t)$ in the complex plane via the expansion \cite{stegmann_short_2016},
\begin{equation}\label{eq:Nzero}
	\langle N \rangle_s (t)= \sum_j \frac{s}{s-1-z_j(t)}\, .
\end{equation} 
The positions of $z_j(t)$ at time $t=61/\Gamma$ are shown in Fig.~\ref{fig:fig3}~(a).
There are complex-conjugated pairs of zeros (blue in Fig.~\ref{fig:fig3}).
Their appearance is well known for systems whose tunneling dynamics is correlated by the presence of Coulomb interaction \cite{stegmann_detection_2015,kambly_factorial_2011,kambly_time-dependent_2013} or superconducting correlations \cite{stegmann_short_2016,souto_quench_2017,kleinherbers_revealing_2018}.
In addition, there are real-valued zeros shown as black dots in Fig.~\ref{fig:fig3}. 
Their evolution with time is shown in Fig.~\ref{fig:fig3}~(b). 
The zeros aggregate near $z=-1$, but periodically with time separation $\pi/\Omega$ an additional zero approaches quickly from $z=-\infty$.
Once this additional zero passes the position $z=s-1$, the denominator in Eq.~(\ref{eq:Nzero}) vanishes and $\langle N \rangle_s (t)$ diverges.

If we restrict the summation in Eq.~\eqref{eq:Nzero} to the black zeros, we find an almost stepwise increase of the number of transferred electrons $\langle N\rangle_1(t)$ with step-to-step distance $\pi/\Omega$ (not shown here). This means that the stochastic system under investigation can be decomposed into a deterministic coherent part (black zeros) and a stochastic incoherent one (blue zeros).
Introducing the weighting factor $s^N$ effectively amounts to distilling out the coherent dynamics.

\begin{figure*}[t]
{\includegraphics[width=\columnwidth]{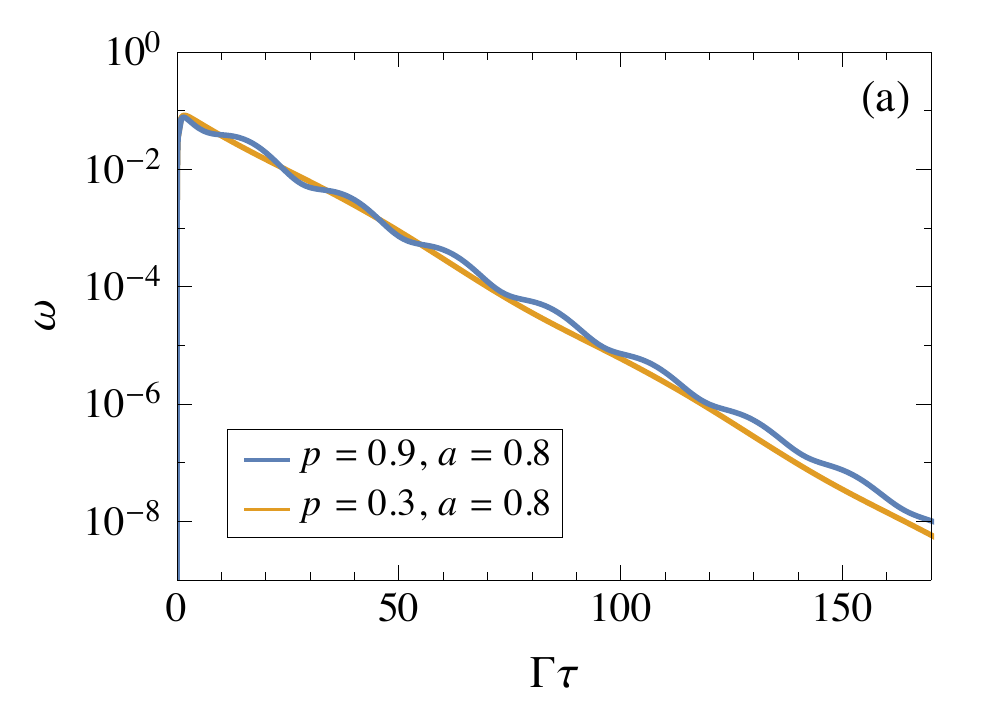}}
{\includegraphics[width=\columnwidth]{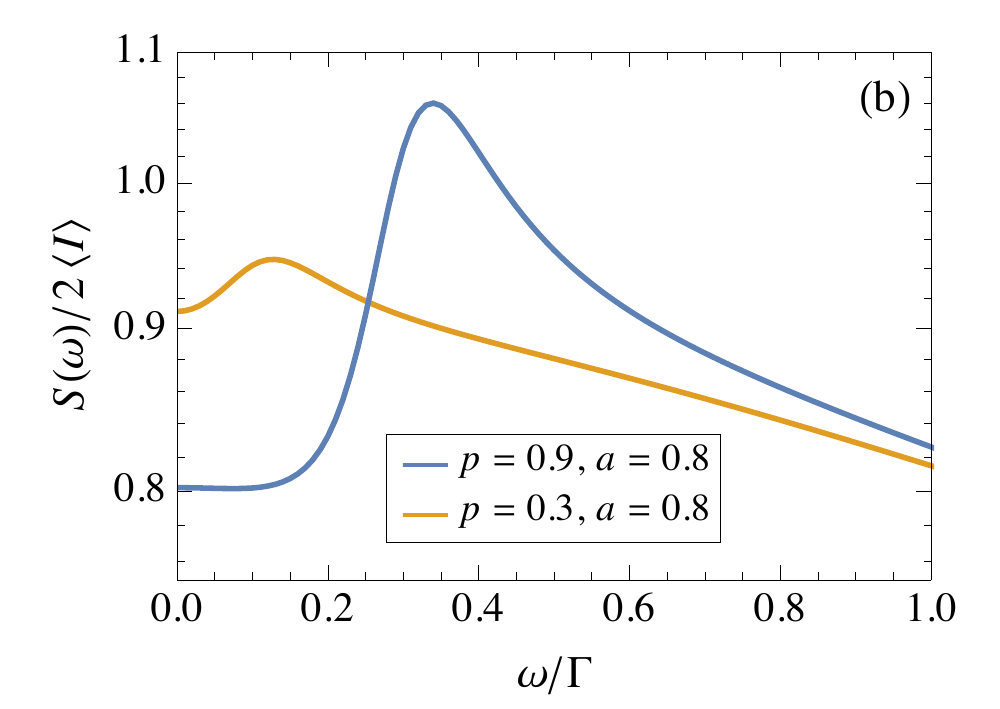}}
\caption{
	(Color online) (a) Waiting-time distribution and (b) finite-frequency Fano factor for two different choices for $p$ and asymmetry $a$, depicted in Fig.~\ref{fig:fig5} by a triangle (blue curves) and a dot (yellow curves). Other parameters are as in Fig.~\ref{fig:fig2}.}
\label{fig:fig4}
\end{figure*}

\begin{figure}[t]
\center
\includegraphics[width=\columnwidth]{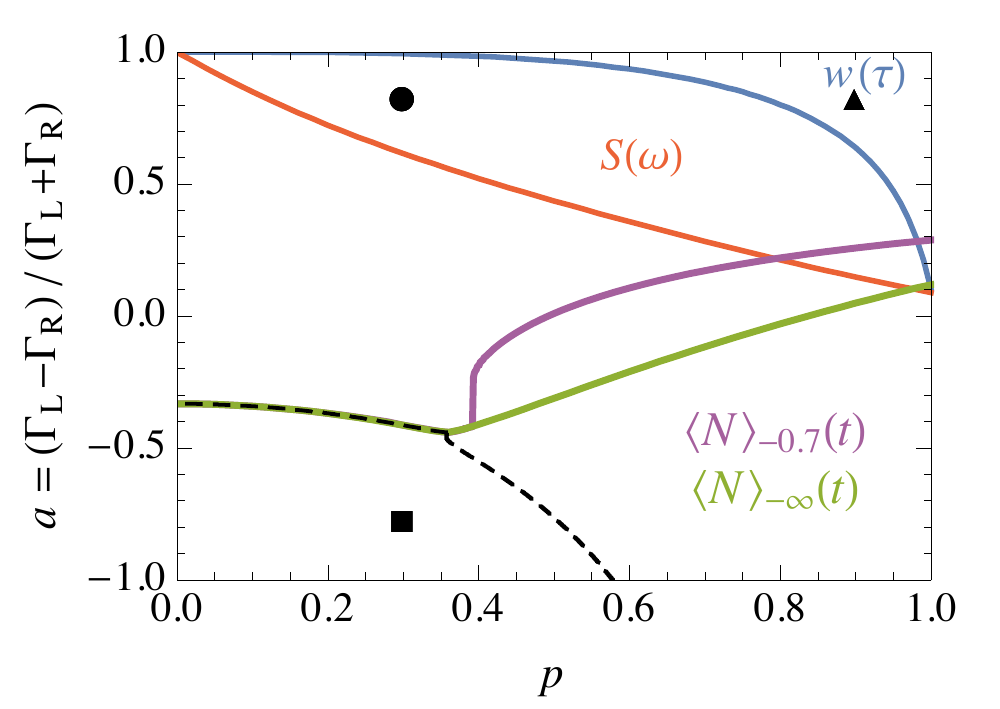}
\caption{\label{fig:fig5} (Color online) Comparison of different detection tools for spin precession as a function of the leads' polarizations $p$ and the asymmetry $a$ of the tunnel couplings to the leads.
Other parameters are as in Fig.~\ref{fig:fig2}. 
Above the blue, red, and green lines, waiting times, Fano factors, and $\langle N\rangle_s(t)$ are useful tools, respectively. 
Obviously, $\langle N\rangle_s(t)$ provides the largest area of application.
Below the dashed black line, $\langle N\rangle_s(t)$ shows divergencies with an $s$-dependent peak-to-peak distance, which are not connected to spin precession. The black dot, square, and triangle marks the values for $p$ and $a$ used in Figs.~\ref{fig:fig4}, \ref{fig:fig6}, and \ref{fig:fig7}.}
\end{figure}

\subsection{Comparison with waiting times and Fano factor}

In order to demonstrate the power of our proposed method, we compare it to two alternative ways to detect coherent spin precession in a quantum-dot spin valve.

First, in the distribution $\w(\tau)$ of waiting times $\tau$ between subsequent tunneling-in and -out events, the precession leads to an oscillation as function of $\tau$, whose presence is deducible via a maximum in the Fourier decomposition. Such an oscillation is illustrated by the blue line in Fig.~\ref{fig:fig4}~(a).

Second, the current noise indicates the coherent spin precession via a maximum in the finite-frequency Fano factor $S(\omega)/2\langle I \rangle$ as a function of the frequency $\omega$. The blue and yellow curves in Fig.~\ref{fig:fig4}~(b) illustrate this particular maximum.

The comparison of the different detection methods is shown in Fig.~\ref{fig:fig5}. Coherent spin dynamics dominates over the incoherent parts for large spin polarization $p\approx 1$ and large asymmetry $a\approx 1$ (upper right corner of Fig.~\ref{fig:fig5}).
In this regime, $\Omega^{-1}>\Gamma_{\rm R}$ such that complete spin precessions occur most likely before a phase-destroying, incoherent tunneling event happens. With decreasing $p$ and $a$, typically incoherent tunneling destroys the  coherences before the spin precession by an angle $\pi$. As a result, the waiting-time distribution can detect the spin precessions only for extreme values of $p$ and $a$ (above the blue line). The finite-frequency Fano factor is somewhat more robust (above the red line in Fig.~\ref{fig:fig5}).
Remarkably, the area in which $\langle N\rangle_s(t)$ displays periodic, $s$-independent divergencies, is much larger, extending to regions in which the dwell time of the electrons is much smaller than $\Omega^{-1}$. For $s=-0.7$ and $-\infty$ the generalized average particle number detects the spin precessions above the purple and green line in Fig.~\ref{fig:fig5}, respectively.
Below $a\lesssim -0.5$, the coherent spin precession is suppressed due to the decoherence introduced by the tunnel coupling to the right lead.
This is modeled by the entry $-\Gamma_{\rm R}$ in the fourth and fifth diagonal matrix elements of $\vec{W}_{\!z}$ given in Eq.~(\ref{Wz}).
Once the coherent spin precession is suppressed, there is nothing left to be distilled out by introducing the weighting factor.

Finally, we remark that the appearance of a divergency in $|\langle N\rangle_s(t)|$ is not always connected to coherent spin precession.
Below the dashed black line in Fig.~\ref{fig:fig5}, coherent spin precession does not play any role for the transport of electrons, as can be seen by inspecting the eigenvalue of $\mathbf{W}_{\!z}$ with the largest real part.
Nevertheless, $\langle N\rangle_s(t)$ exhibit divergencies periodic in time (see~Appendix~B).
Their peak-to peak distance is, however, {\it not}~determined by the Larmor frequency $\Abs{\mathbf{B}}$. 
In contrast to the divergencies shown in Fig.~\ref{fig:fig2}, the peak-to-peak distance strongly depends on $s$ (it scales with $\sqrt{\Abs{s}}$), and it is independent of $\varepsilon$, $U$, and $\mu_r$.

\subsection{Statistical accuracy}

\label{sec:req}
\begin{figure}[t]
\center
\includegraphics[width=\columnwidth]{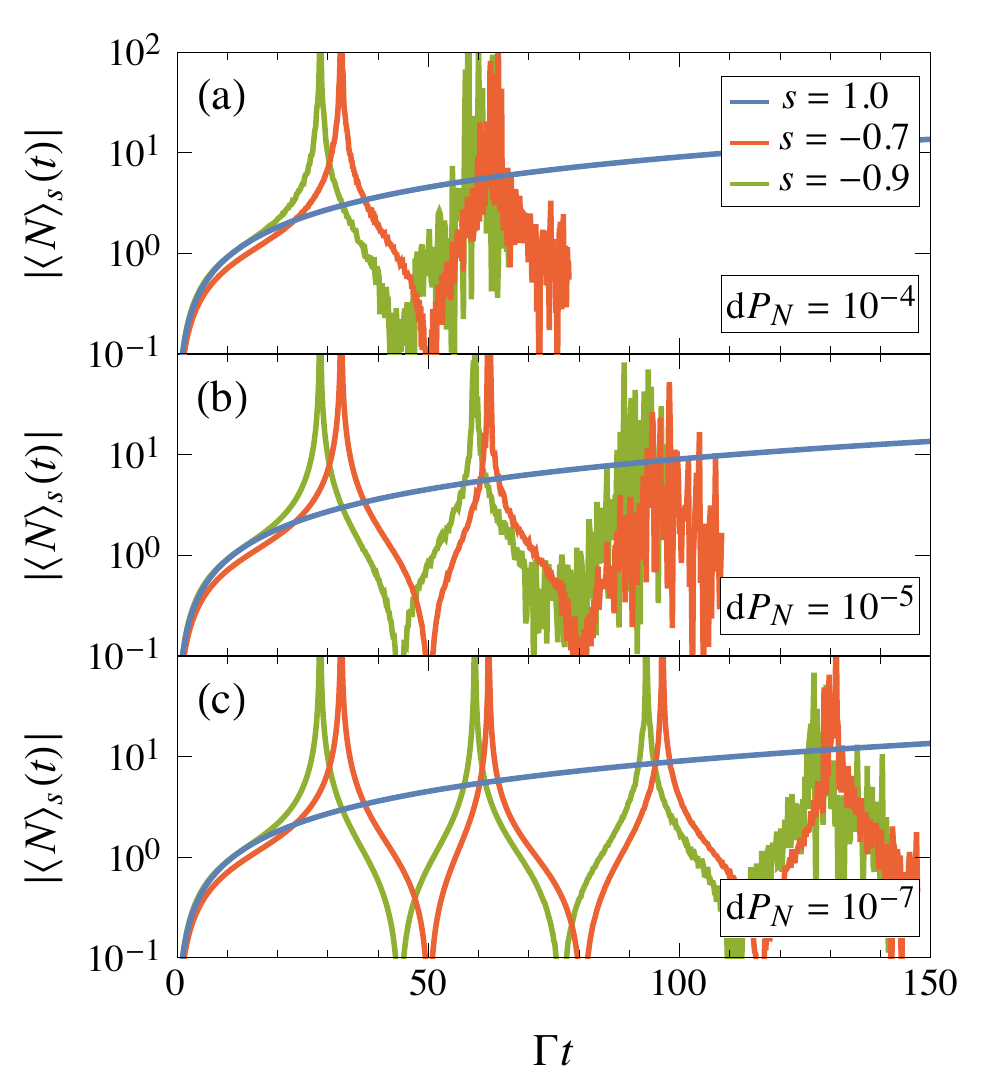}
\caption{\label{fig:fig6} (Color online) $|\langle N\rangle_s(t)|$ as a function of time for $s=1$ (blue curve), $-0.7$ (red curve), and $-0.9$ (green curve). From (a) to (c), the precision $\tx{d}P_N$ of $P_N(t)$ increases and more peaks are resolved. The parameters are as in Fig.~\ref{fig:fig2}, with polarization $p$ and asymmetry $a$ as for the dot in Fig.~\ref{fig:fig5}.
}
\end{figure}

The formula for $\langle N\rangle_s(t)$, Eq.~(\ref{genN}), contains a series over $N$ of the probability distribution $P_N(t)$.
Therefore, the accuracy of $\langle N\rangle_s(t)$ calculated from experimentally measured data is limited by the finite measurement time (which cuts the infinite series into a finite sum) and the precision with which the probability distribution $P_N(t)$ can be experimentally determined. 
In this section, we discuss how the value of $s$ and, thus, the weighting factor $s^N$ affects this accuracy. 

For $|s|>1$, the weighting factor favors contributions with larger $N$. 
The convergence of the series is guaranteed as long as $P_N$ falls off fast enough with $N$. 
The probabilities of a Poisson distribution, for example, contain a prefactor $1/N!$ that guarantees convergence, and the upper cutoff at which the series can be practically terminated only grows linearly with $|s|$, which does not pose a serious challenge for real experiments.

A more severe issue is the precision of the measured probability distribution $P_N(t)$.
With increasing $|s|$, the value of $\langle N\rangle_s(t)$ is dominated by large-$N$ probabilities $P_N$ that are small and, therefore, difficult to resolve, which, in turn, reduces the accuracy of $\langle N\rangle_s(t)$.
The most accurate results are, therefore, expected for small $|s|$.
For $s=0$, on the other hand, the divergencies in $\langle N\rangle_s(t)$ that indicate the spin precession are gone.
This motivates the choices $s=-0.7$ and $-0.9$ in our calculation. 

To estimate the required precision of $P_N$ to resolve several divergencies of $\langle N\rangle_s(t)$, we perform the following simulation. 
After calculating $P_N(t)=\partial^N_z \mathcal{M}_0(z,t)|_{z=0} /N!$ from the moment-generating function $\mathcal{M}_0(z,t)$, we artificially introduce an error by rounding the obtained $P_N$ to the nearest multiple of a chosen precision $\tx{d}P_N$.
The result for $\langle N\rangle_s(t)$ for parameters as in Fig.~\ref{fig:fig2} is shown in Fig.~\ref{fig:fig6}.
We see that by improving the precision $\tx{d}P_N$, more and more divergencies can be resolved.

% - - - - - - - - - - - - - - - - - - - - - - - - - - - - - - - - - - - - - -
\section{Conclusions}
We propose a new method to detect coherent dynamics in stochastic processes, particularly the coherent spin precession in a quantum-dot spin valve.
Full counting statistics of the number of transferred electrons as function of time is utilized to distill out the coherent part out of the statistics that is predominantly probabilistic incoherent in nature. 
The key idea is to introduce a weighting factor $s^N$ when calculating the average number $\langle N\rangle_s(t)$ of transferred electrons.
For $s<0$, coherent precession due to an exchange field is detectable by a periodic appearance of divergencies in $\langle N\rangle_s(t)$.
The peak-to-peak distance $\pi/\Omega$ is approximatively determined by the Larmor frequency, $\Omega \approx |\mathbf{B}|$.
Our proposal works for wide range of polarizations and asymmetries of the tunnel couplings. 
In particular, it allows to use weakly polarized ferromagnets such as Ni alloys and can be applied even if alternative tools such as finite-frequency current noise and waiting-time distribution fail.
An experimental proof of the concept seems well in reach with recent experimental setups.

% - - - - - - - - - - - - - - - - - - - - - - - - - - - - - - - - - - - - - -
\section{Acknowledgements}
We acknowledge financial support of the Deutsche Forschungsgemeinschaft (DFG) under Projects No. WE $5733/1-2$, No. KO $1987/5-2$, and No. SFB 1242 TP~A2. We also thank B. Sothmann for valuable discussions. 

\onecolumngrid
% - - - - - - - - - - - - - - - - - - - - - - - - - - - - - - - - - - - - - -
\appendix
\section{Explicit form of the generator $\vec{W}_{\! z}$} 
As outlined in the main text, the generalized-factorial-moment-generating function $\mathcal{M}_s(z,t)$ can be calculated from the generator $\vec{W}_{\! z}$.
To write down the latter explicitly, we choose the spin quantization axis along the magnetization direction of the left (source) electrode [any other choice leads to a different $\vec{W}_{\! z}$ but, of course, results in the same $\mathcal{M}_s(z,t)$].
We find
\begin{equation}
	\label{Wz}
	\vec{W}_{\!z} =\begin{pmatrix}	
	 -2\Gm_{\tL}& z (1+p \cos{\phi})\Gm_{\tR} & z (1-p \cos{\phi})\Gm_{\tR} & i z  p \Gm_{\tR} \sin \phi &  -i z p \Gm_{\tR} \sin \phi  \\
	(1+p)\Gm_{\tL} & -(1+p \cos{\phi})\Gm_{\tR}&  0 & \frac{B_{\tR}- i p \Gm_{\tR}}{2}\sin \phi  & \frac{B_{\tR}+ i p \Gm_{\tR}}{2}\sin \phi  \\ 
	(1-p)\Gm_{\tL}&0 & -(1-p \cos{\phi})\Gm_{\tR} & \frac{-B_{\tR}- i p \Gm_{\tR}}{2}\sin \phi  & \frac{-B_{\tR}+ i p \Gm_{\tR}}{2}\sin \phi  \\
	0& \frac{-B_{\tR}+ i p \Gm_{\tR}}{2}\sin \phi &\frac{B_{\tR}+ i p \Gm_{\tR}}{2}\sin \phi& - \Gm_\tR +i (B_{\tL}+  B_{\tR} \cos \phi) & 0 \\
	0& \frac{-B_{\tR}- i p \Gm_{\tR}}{2} \sin \phi & \frac{B_{\tR}- i p \Gm_{\tR}}{2}\sin \phi & 0&  - \Gm_\tR -i (B_{\tL}+B_{\tR} \cos \phi) \\
	  \end{pmatrix}.
\end{equation}
Here, ${B_r} = \frac{p \Gamma_r}{2\pi} \left[\Psi (\epsilon+U-\mu_r) -\Psi (\epsilon-\mu_r) \right]$ is the magnitude of the contribution to the exchange field that is generated  by $r=\tL,\tR$.
The first three columns describe transitions where the initial state is the quantum dot being empty, singly occupied with spin $\uparrow$ , and singly occupied with spin $\downarrow$, respectively.
For the fourth and fifth columns, the initial state is a coherent superposition of spin $\uparrow$ and $\downarrow$.
The fact that the counting field $z$ appears only in the first row indicates that we count those tunneling events where an electron is leaving the quantum dot to the right lead.

\section{Estimate of the peak-to-peak distance of the divergencies} 

To estimate the period with which the divergencies of $\langle N\rangle_s(t)$ appear, we analyze the eigenvalues of $\mathbf{W}_{\!z}$.
In order to get compact and transparent explicit expressions, we concentrate on the limit of large $|s|$, i.e., we expand the eigenvalues in orders of $1/|z|$.
The leading contributions are
\begin{align}
	\lambda_{1}&=-\Gm_\tx{R}+\frac{\Gm_\tx{R}  \left(	B_{\tL}	+	B_{\tR}	\right)	\left(	B_{\tL} \cos \phi	+	B_{\tR}	\right) (1+\cos \phi)}{\Abs{\mathbf B}^2} p^2	+ \mathcal{O}(p^4)  , \\
	\lambda_{2,3}&=-\Gm_\tx{R}		+	\frac{\Gm_\tx{R}	B_{\tL}	\left(	B_{\tL}	-	B_{\tR}	\right)	\sin^2 \phi}{2 \Abs{\mathbf B}^2}p^2 	\pm 	i \Abs{\mathbf B} + \mathcal{O}(p^3), \\
	\lambda_{4,5}&=-\Gm_\tx{L}-\Gm_\tx{R}	+	\frac{(1-p^2)\Gm_\tx{R}}{2\left(1+p^2 \cos	\phi \right)} \pm 	i  \sqrt{-2\Gm_\tx{L}\Gm_\tx{R}z \left( 1+p^2 \cos  \phi \right)}		 \, . 
\end{align}
\twocolumngrid
The first eigenvalue is purely real while the others come (for negative $z$) as complex-conjugated pairs. The real part of all of the eigenvalues is negative.

For times $t\gg 1/\Gamma$, the dynamics of the system is determined by only the eigenvalue $\lambda_\tx{max}$ of $\mathbf{W}_{\!z}$ with the largest real part (i.e., the one closest to $0$). 
In the region above the green line in Fig.~\ref{fig:fig5}, the dominating eigenvalues are $\lambda_2$ and $\lambda_3$. 
Their imaginary part is given by the Larmor frequency $\Abs{\mathbf B}$, independent of $z$. 
This explains the periodicity of the divergencies of $\langle N\rangle_s(t)$.

\begin{figure}[h]
\center
\includegraphics[width=1\columnwidth]{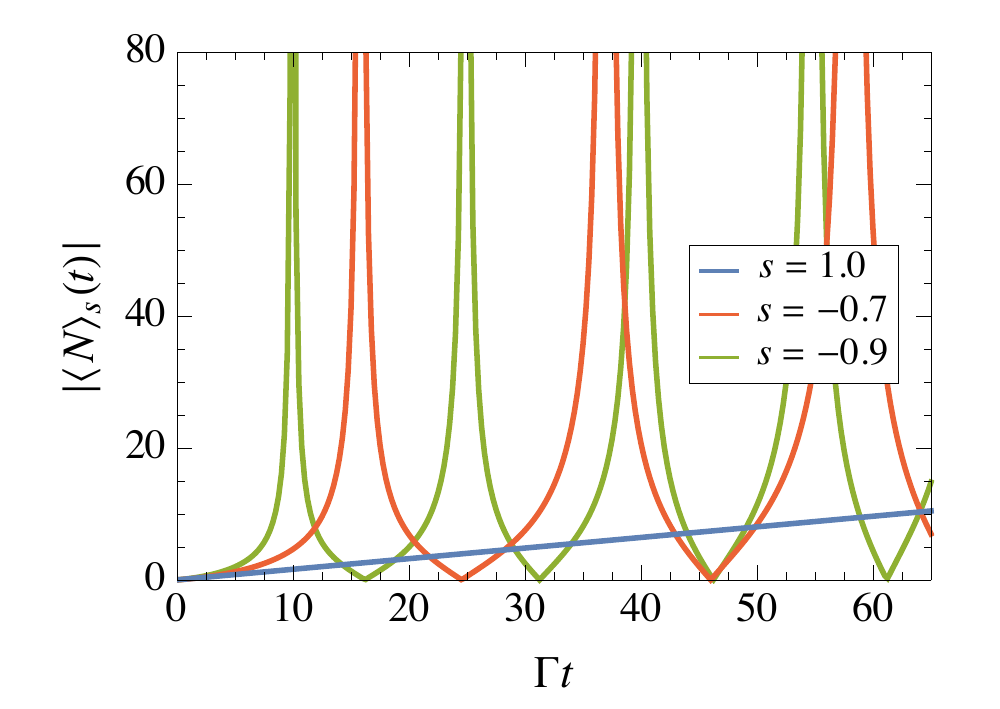}
\caption{\label{fig:fig7} (Color online) $|\langle N\rangle_s(t)|$ as a function of time for parameters $a=-0.8$ and $p=0.3$ (corresponding to the square in Fig.~\ref{fig:fig5}). The other parameters are as in Fig.~\ref{fig:fig2} of the main text.}
\end{figure}

In the region below the black dashed line in Fig.~\ref{fig:fig5}, the dominating eigenvalues are $\lambda_4$ and $\lambda_5$. 
They also have some finite imaginary part, but this time we get oscillations with frequency $\sqrt{-2 \Gm_\tx{L} \Gm_\tx{R}s \left(1+p^2 \cos \phi \right)} $ that are not associated with Larmor precession.
In particular, this frequency depends on $s$, as depicted in Fig.~\ref{fig:fig7}.

% - - - - - - - - - - - - - - - - - - - - - - - - - - - - - - - - - - - - - -

\end{document}